# An Approach for Self-Adaptive Path Loss Modeling for Positioning in Underground Environments

Evgeny Osipov, evgeny.osipov@ltu.se 971 87, Sweden, Luleå, Luleå University of Technology, the Department of Computer Science Electrical and Space Engineering

Denis Kleyko, denis.kleyko@ltu.se 971 87, Sweden, Luleå, Luleå University of Technology, the Department of Computer Science Electrical and Space Engineering

Alexey Shapin, alexshapin@gmail.com 630102, Russia, Novosibirsk, Siberian State University of Telecommunications and Information Sciences, the Department of Automatic Electrical Communication

**This paper proposes a real-time self-adaptive approach for accurate path loss estimation in underground mines or tunnels based on signal strength measurements from heterogeneous radio communication technologies. The proposed model features simplicity of implementation. The methodology is validated in simulations as well as verified by measurements taken in real environments. The proposed method leverages accuracy of positioning matching to the existing approaches while requiring smaller engineering efforts.**

## Introduction

Accurate positioning in underground environments like tunnels or mines is essential for their efficient operation (e.g., logistics, energy savings, improved decision support) as well as for personal safety e.g., reducing the number of incidents. For example in the mining industry there is a need for positioning of people, equipment and vehicles. At the same time precise positioning based on the properties of the radio signal propagation requires accurate characterization of the radio channel, which is challenging to generalize for a given technology, due to the unique geometry of the corridors as well as the reflective properties of the walls' material. This article elaborates on the feasibility of designing a positioning service built upon the existing standard wireless infrastructure deployed in mines. Wireless communication networks became an integral part of the mining process nowadays: Several large mines e.g., in Canada [1], Sweden [2], and Bulgaria [3] have deployed underground IEEE 802.11 based networks covering all tunnels. However, Wi-Fi systems alone cannot be considered for accurate positioning. Therefore the main motivation for this work is that other radio systems with higher accuracy in combination with IEEE 802.11 networks should jointly enhance the accuracy of positioning. An example of such a complementary system is RFID with passive tags, which is often installed for safety monitoring, monitoring of assets and environmental conditions, controlling of delivery cycles of trucks, collision avoidance between vehicles in tunnels, etc. Several companies provide and deploy RFID solutions worldwide (see [19] for more details on the RFID deployment in Australia, Chile, India and Sweden). This article presents a methodology, which utilizes the diversity of the deployed radio technologies and provides a robust and accurate positioning service by combining information from multiple heterogeneous wireless technologies.

The traditional methods of communication system-aided positioning include time-of-arrival, angle-of-arrival and received signal strength indication (RSSI) approaches [4]. The key aspect in all these approaches is the transformation of the signal strength level into the actual distances. The decisive factor influencing the accuracy at this step is the quality of the adopted radio-propagation model and the ability to accurately estimate the radio signal path loss in the given environment. While in most of outdoor scenarios the simple free-space path loss model (FSPL) [5] provides sufficient accuracy, positioning process in underground scenarios is challenging due to complex electromagnetic characteristics of the environment. In underground environment the signal propagation characteristics depend on many factors e.g., the surface of a tunnel, the geometry of the tunnel, the type of a positioning object (a car, a truck or a human), the presence of obstacles between the receiver and the transmitter, the presence of mobile objects etc.

Practically it is known that that there is no good enough analytical model for describing all possible variations of the physical topologies. This article proposes a practical approach for self-adaptive estimation of the signal path loss in the given deployment site. The main contribution is formulated as a method for self-adaptive template path loss model based on aggregation of data from heterogeneous wireless technologies.

The article is organized as follows. An overview of the related work is provided in Section 2. Section 3 describes the system-level architecture. Section 4 presents self-adaptive iterative path loss model. Section 5 provides validation of the proposed methodology and its real-life performance. Finally, Section 6 discusses and concludes the article.

**Related work**

The current state of the art of wireless communications and propagation modelling in underground mines can be found in [6, 7]. The existing methods can be classified as: Numerical methods for solving Maxwell equations [8, 9]; Waveguide based channel models [10, 11]; Ray tracing based channel models [12]; and Empirical path loss modelling. The approach presented in this article falls in the latter domain. The main models are so called *N*-slopes models, where the signal attenuation is modelled differently on *N* discrete intervals from the source of the radio signal. For instance [13] proposes a four-slope path loss model for car tunnels. The work in [14] further extends the model into a five-zone model for the scenario when the size of object is comparable with the size of the tunnel and the transmission range could be up to several thousand meters. For this article we adopt the two-piece path loss model [15], which features computational simplicity while being known for adequate accuracy. We argue that even the simplest model can leverage competitive performance when supported by information from several communication technologies.

**System-level architecture description**

Consider a generic communication system to be deployed in an underground environment as illustrated in Figure 1. The system deploys heterogeneous wireless technologies: an IEEE 802.11 based infrastructure network and a near field communication (NFC) system. The IEEE 802.11 based network consists of base stations (BS) and transceivers installed on different mobile and static objects e.g., vehicles, workers, facilities, etc. The near field communication system consists of RFID tags and RFID readers. In mines the NFC is used for various purposes including access control and inventory of objects. The coordinates of both IEEE 802.11 based BSs and RFID readers are known exactly.

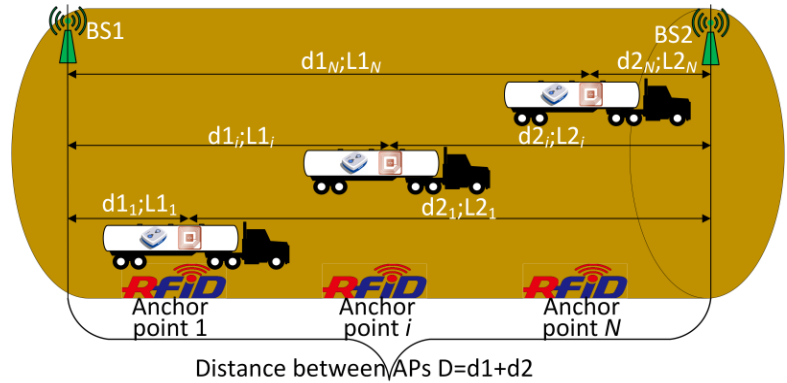

*Figure 1* System model.

Note that in reality several wireless networks (also from different vendors) could be deployed on orthogonal channels for serving different purposes for example a network for VoIP and personal data communications, a network for remote control of automatic vehicles etc. This article, however, considers a generic case of a single production network with virtualization capabilities connected to the main control room. Thus, it is assumed that all base stations are interconnected by a wire line network forming an Extended Service Set. It is also assumed that the RFID readers are connected to the same production network either wirelessly or directly. Therefore, all information from all connected devices is available centrally.

The client side of the system is represented by RFID tags and different kind of IEEE 802.11 based devices including data communication units (personal communicators, devices for remote control of vehicles, etc.) as well as Wi-Fi based asset tags [16]. The latter devices are specifically designed for Wi-Fi based positioning. The number of deployed RFID readers between two base stations is $N_{RFID} \in [0 \ldots N]$ (anchor points): While some tunnels could be equipped with several RFID readers, other tunnels might lack any at all.

*Outline of the solution*

The proposed solution addresses one of the fundamental challenges of accurate modelling of radio signal path loss in the underground environment. The main idea is to adopt the simplest possible model of signal path loss with distance (two-piece model in Figure 2) with the initially unknown parameters. Despite its simplicity the model captures the major radio propagation modes. In the near region, the path loss slope is steep and in many cases it is modelled as free space path loss. In the far region, the waveguide effect appears with few lower order modes and the path loss slope is reduced significantly [7]. This model will be used as a *template* for the target model to be used for RSSI-based positioning at Wi-Fi base stations. Note that while fading is an important effect for wireless propagation models it is not accounted by the adopted two-piece template. The aim of the proposed methodology is to study the feasibility of accurate positioning with a simple model in a real-life underground environment. Overall the goal is formulated as to come up with accurate and *tunnel-specific* estimates of the path loss by overlaying the events of measuring the signal strength from a particular node at the base station with the events when this monitoring node is engaged with one of the RFID readers. This will be done by constructing and solving a system of $2 \cdot N_{RFID}$ equations describing the model and the corresponding signal strength measurements $L_i$ Figure 1 at points of engagement with RFID readers ($d_i$ in Figure 1). The system of equations will be solved iteratively, where the iteration is a passage of an object through all RFID readers in the particular tunnel.

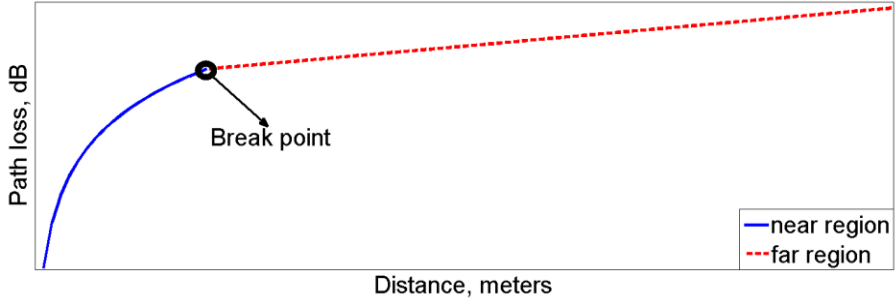

*Figure 2* Two-piece model.

The resulting model could be used then in two ways. Firstly, in tunnels featuring RFID readers the model will be continuously updated accounting for dynamically changing properties of the propagation environment due to mobility of the monitored nodes. In this case the accuracy of the estimated model will be the highest one. Secondly, the averaged model will be used statically in corridors lacking RFID technology, but having similar geometry. In this case the accuracy of the model will be lower than in the previous case, however, it will be higher compared to the case when model's parameters are chosen according to commonly used assumptions without anchoring to the specifics of the particular operation environment.

**Self-adaptive path loss model**

This section presents the details of the proposed methodology. First, the iterative model construction is presented. Then the operation phase is discussed.

*Iterative model*

As argued earlier for simplicity of the engineering the two-piece model [7] is adopted. The two-piece model defines two (*near* and *far*) regions. The path loss in the *near* region is modelled by the free-space path loss model (1), where $\gamma$ is the path loss exponent, $C$ is the constant, which takes into account system losses and $d$ is the distance between the transmitter and the receiver. $L$ is the path loss in dB. The waveguide effect, which reduces the path loss, is captured by the model of the *far* region. The path loss in the *far* region is modelled by a linear approximation (2) [13], where $\alpha$ is the curve slope, $L_0$ and $d_0$ are the path loss and distance at the break point respectively.

$$L[\text{dB}] = \gamma \cdot (10 \log_{10}(d) + C) \tag{1}$$

$$L[\text{dB}] = L_0[\text{dB}] + \alpha \cdot (d - d_0) \tag{2}$$

The break point $d_0$ divides the whole distance into the two regions and can be estimated from a free first Fresnel zone (3) [5], where $h_r$ and $h_t$ are heights above the ground, $\lambda$ is the wavelength. For the proposed approach the break point position is the adjustable parameter.

$$d_0 = \frac{4 h_r h_t}{\lambda} \tag{3}$$

Summarizing the above discussion the two-piece model, which we further on refer to as the *template model* is presented in (4).

$$\begin{cases} L[\text{dB}] = \gamma \cdot (10 \log_{10}(d) + C), & \text{when } d \leq d_0 \\ L[\text{dB}] = L_0[\text{dB}] + \alpha \cdot (d - d_0), & \text{otherwise} \\ L_0[\text{dB}] = \gamma \cdot (10 \log_{10}(d_0) + C) \end{cases} \tag{4}$$

The template model has four unknown parameters, namely: $\gamma$, $C$, $d_0$ and $\alpha$. These parameters will be estimated and tuned iteratively upon collecting the data from heterogeneous wireless sources. Now, each time the RFID and the Wi-Fi enabled node passes by an RFID reader it also communicates with neighbouring base stations.

According to the system model in Figure 1 the object is located between two BSs, distance $D$ between BSs is known. When the object passes by the $i$th RFID reader its distances to base station 1 ($d1_i$) and base station 2 ($d2_i$) are determined. At the same time BSs measure RSSI of the ping pulses issued by the object, i.e. path losses $L1_i$, $L2_i$ are known as well. Thus, the system of two equations (5) is constructed for each RFID reader, where $i$ is the $i$th reader.

$$\begin{cases} L1_i[\text{dB}] = \gamma \cdot (10\log_{10}(d1_i) + C), & \text{when} \quad d1_i \leq d_0 \\ L1_i[\text{dB}] = L_0[\text{dB}] + \alpha \cdot (d1_i - d_0), & \text{otherwise} \\ L2_i[\text{dB}] = \gamma \cdot (10\log_{10}(d2_i) + C), & \text{when} \quad d2_i \leq d_0 \\ L2_i[\text{dB}] = L_0[\text{dB}] + \alpha \cdot (d2_i - d_0), & \text{otherwise} \end{cases} \qquad (5)$$

During the runtime a system of up to $N_{RFID}$ systems of equations of form (5) are constructed at each iteration. The system is then solved numerically e.g., using nonlinear least-squares method [17].

*Operating phase*

When the four parameters for the template model are estimated the distance from an object to the BS can be calculated as in (6).

$$\begin{cases} d = 10^{\frac{L[\text{dB}]}{\gamma} - C}, & \text{when} \quad L[\text{dB}] \leq L_0[\text{dB}] \\ d = d_0 + \dfrac{L[\text{dB}] - L_0[\text{dB}]}{\alpha}, & \text{otherwise} \end{cases} \qquad (6)$$

Note that when the object is located between two base stations as in Figure 1 the additional condition such as the distance between base stations $D$ should be considered. Obviously, due to errors in the estimation of $d1$ and $d2$ their sum may be lesser or larger, then the distance $D$. Therefore, the obtained values are normalized on $D$ with respect to (7). One should consider that normalization is appropriate only if the object is positioned between base stations. Moreover, the estimated model will work even in the case when there is a signal level only from one BS i.e., the normalization is impossible.

$$\begin{cases} \tilde{d1} = D \cdot \dfrac{d1}{d1 + d2}, & \text{when} \quad d1 + d2 > D \quad \text{or} \\ \tilde{d2} = D \cdot \dfrac{d2}{d1 + d2}, & \text{when} \quad d1 + d2 < D \end{cases} \qquad (7)$$

Thus for the system in Figure 1 in order to form the model and subsequently estimate the distances the following actions are needed:
- Choose the template model (4);
- Detect that the object has passed $i$th RFID reader;
- Calculate the distances $d1_i$ and $d2_i$ from the reader to the base stations;
- Measure path losses $L1_i$ and $L2_i$ of the object's signal at the base stations;
- Construct a system of equations in the form (5);
- Repeat the above steps for $N_{RFID}$ readers in the tunnel;
- Estimate the parameters for the template model by solving the system of equations e.g., using nonlinear least-squares method [17];
- Calculate the distances $d1$ and $d2$ from the object to the bases stations according to (6);
- Whether it is applicable normalize the distances as in (7).

**Validation of the proposed methodology and its real-life performance**

For the validation of the proposed methodology of the signal path loss and its convergence properties the following approach was adopted. The two-piece model with arbitrary chosen parameters is calculated for stretch [0..300] meters and used as the reference case. The reference model's parameters were set as $\gamma = 2$, $C=20.1$, $d_0=50$ meters and $\alpha = 0.2$. Note that this choice of parameters' values does not correspond to any particular operating environment. While the model for the validation might at the first glance seem unjustifiably similar to the template model it is a valid choice for the pragmatic validation approach. That is to check first whether the proposed approach is viable for at least an artificial model with similar structure and known parameters. Next the assessment of the approach in real-life environments is presented in the subsequent subsections.

The training set for the proposed iterative method is constructed by first sampling the calculated values of the two-piece reference model in *N* anchor points (*N* will be varied in the simulations). In order to reflect the fact that real RSSI values are noisy the additive white Gaussian noise was randomly added to the each sampled value resulting in SNR value being -2 dB on average, the standard deviation of the noise was 1.25 dB. This is reflected by the red curve in Figure 3. Note that SNR value was chosen to be low in order to demonstrate that the parameters for the template model can be estimated even in noisy environment.

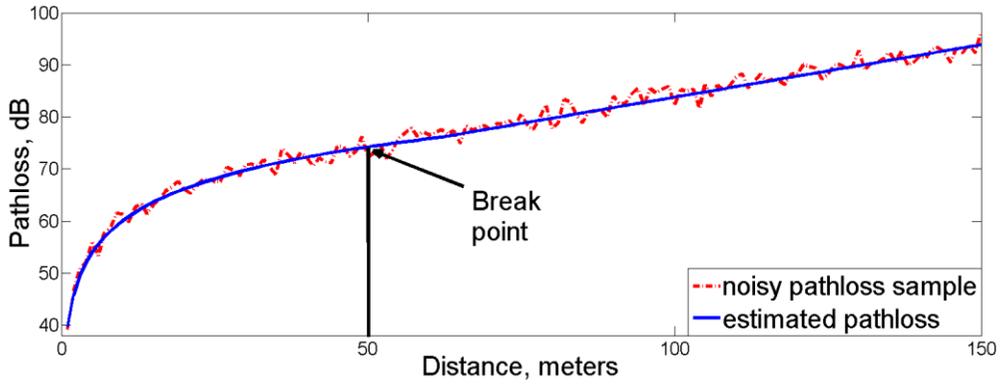

*Figure 3 Path loss of noisy signal sample and estimations made by the proposed method using four non-uniformly distributed anchor points.*

The task of validation is formulated as to apply the proposed method on the noisy sequence of RSSI values, then to estimate the parameters for the template model and finally compare the estimated values to the parameters of the reference model.

The model parameters were estimated in four simulations for different number of anchor points $N \in [19, 14, 9, 4]$. In the simulations with *N*=19, *N*=14 and *N*=9 the distance between points were distributed uniformly on a stretch of 300m. The simulation with *N*=4 was conducted with the non-uniform placement of the anchor points by reason described in below. In this simulation the anchor points were set on 15, 30, 270, 285 meters.

Each simulation for a given number of anchor points consisted of 100 iterations. At each iteration one RSSI sample with *N* values was generated. At each iteration the samples from the previous iterations were added to the training set and the average value of path loss $\bar{L}$ was calculated for each anchor point. This is done in order to understand the accuracy and the convergence time properties of the proposed self-adaptive path loss estimation methodology.

Figure 3 shows an example of results obtained from the iteration of a simulation with 100 samples for four anchor points. The red dashed curve shows one of noisy samples used in this iteration and the solid curve shows the two-piece model computed with the estimated by the proposed method parameters.

*Convergence and the factors affecting the accuracy*

Figure 4 illustrates the development of the four estimated parameters with the increasing number of samples taken into consideration and different number of anchor points considered in the simulation. The main observation concerns the convergence property of the proposed estimation method. It is easy to see that all four estimated parameters converge differently to the values of the reference model for different number of the anchor points. While with 19 anchors the estimated by the proposed self-adaptive method parameters (the red curves in the figure) converge to the target values (the straight line in the figure) fairly quickly, the convergence time worsens when the number of anchors is decreasing. With 9 anchors the proposed method does not converge to the reference values. This is because with uniform distribution of the small numbers of anchors over the monitored stretch most of them fall into the *far* region of the two-piece model. With 19 anchors distributed evenly over the same stretch the number of nodes in both the *near* and the *far* regions are sufficient for the estimation. The large number of anchors is, however, not practical to have from the economical point of view.

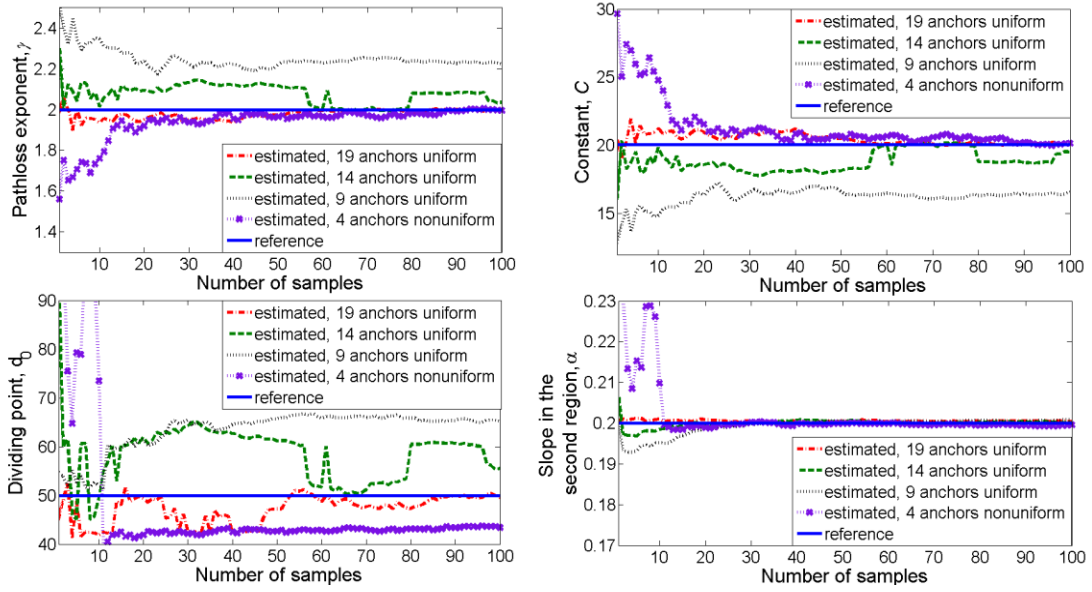

*Figure 4 Dynamics of the estimated parameters.*

We, therefore, simulated a case with the non-uniform placement of the anchor points. Specifically, the points in the four anchors simulation were set on 15, 30, 270, 285 meters. With this placement two nodes were placed in the presumed *near* region of the base station at the beginning of the stretch and other two anchors were placed in the presumed *near* region of the base station at the far end of the stretch. For any of the two base stations the other two nodes were located in their corresponding *far* region. The results of the simulations (shown by the magenta coloured curves in Figure 4) show a very good convergence of all estimated parameters to the reference values.

This observation gives an important insight onto the rules for the placement of the RFID readers in the underground environment: If the RFID readers are intended to be used for calibration of the Wi-Fi based positioning system, their limited number should be placed in a relative proximity to the base station, i.e. in the presumed *near* region of the transmitter. Observed results promise that the approach will also be viable in a real-life underground environment.

*Assessment of the estimation accuracy in real environments*

The accuracy of the proposed self-adaptive path loss estimation model was assessed in two real-life operating environments. The first environment is an 80 meters underground corridor in the university building. The second environment is two 300 meters long tunnels in an ore mine in northern Sweden. The following subsections report the results of these experiments.

*Method's accuracy in the underground corridor*

The length of the considered corridor is approximately 80 meters. The measurements were taken using three laptops. Two of them were located at distances 0 and 80 meters and played roles of base stations. The third laptop was used for imitation of the moving object. The RSSI values from the base station were taken with 2 meters interval. Thus measurements from 38 points were used to obtain the estimate of the path loss model parameters. Note that the 38 locations here are used for testing purposes only and this number does not indicate the number of needed RFID readers. On contrary the results of the validation show that only few readers are sufficient for the estimation of parameters. The measurements and the result of the estimation are depicted in Figure 5. The red circles in the figure denoted as $RSSI_{BS1}$ in the legend correspond to the measurements taken at the base station on the one end of the corridor and blue crosses, denoted as $RSSI_{BS2}$, reflect the RSSI values measured by the base station at the other end of the corridor.

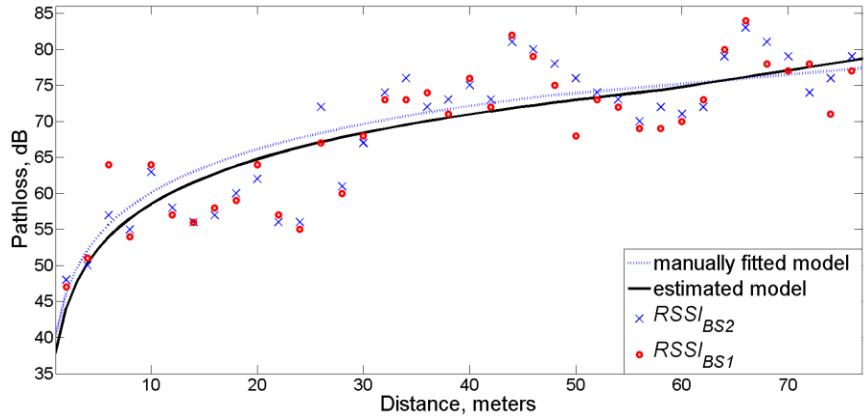

*Figure 5* Comparison between hand calculated and estimated models.

In order to assess the performance of the proposed method a reference model was calculated using the parameters of FSPL model in the near region, the break point was calculated from 3) and the slope for the far region was fitted based on the measured values. This reference model is illustrated by the blue dotted curve in Figure 5. The model based on the estimates of the parameters from the proposed iterative model is shown by the solid black curve in the figure. The graph shows no significant difference between the model based on the estimated by the proposed methodology model and the best one manually fitted to the measurements.

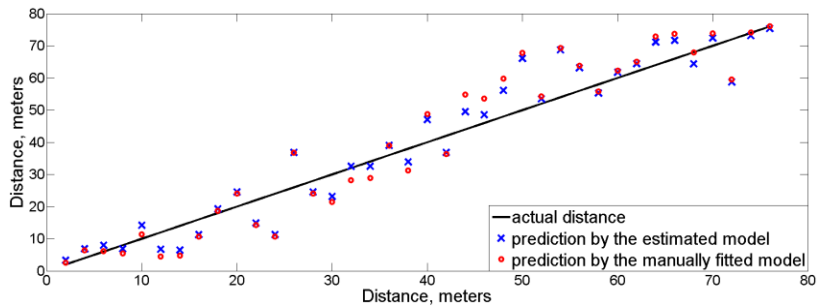

*Figure 6* Predicted distances by hand calculated and estimated models.

Note that in the operating phase the proposed model is used for the estimation of the distance to the BS. Hence it is important that the distance predictions by the proposed model are similar to predictions from the hand-calculated model. The actual distances and the results of prediction by the models normalized with (7) are depicted in Figure 6. The actual distances are shown by the solid black curve in the figure. The blue crosses in Figure 6 correspond to the distance predictions made by the model with parameters estimated by the proposed approach. The red circles reflect the distance predictions made by the reference manually fitted model. These results show that the estimated model and the reference model have the same level of accuracy to the measurements. However, there is no need to do manual calculations and know the deployment specific in the environment for the estimation of model parameters with the proposed approach.

*Method's accuracy assessed in the ore mine*

This time the model was assessed in Kristineberg ore mine [18] located in northern Sweden. The mine is equipped with a IEEE 802.11-based network from Cisco operating in 2.4 GHz band. On the testing site most of the mine's corridors are covered by Wi-Fi network. Wi-Fi access points are deployed such that the objects are visible by at least two access points (see a segment of the mine in Figure 7). Thus, communications are mainly line-of-sight except for some tunnels, which are curved. The training set was obtained by recording RSSI level in ping pulses issued by Ekahau asset tags [16]. Figure 7 shows the layout of mine's areas, where the measurements were taken.

During the measurements the tunnels were empty. Since the mine currently is not equipped with RFID readers, measurements were taken manually similarly to the corridor scenario. The distances between anchor points were uniformly distributed with 15 meters intervals. RSSI values on base stations were stored and associated with the anchor points. All anchor points were used for the template path loss model estimation.

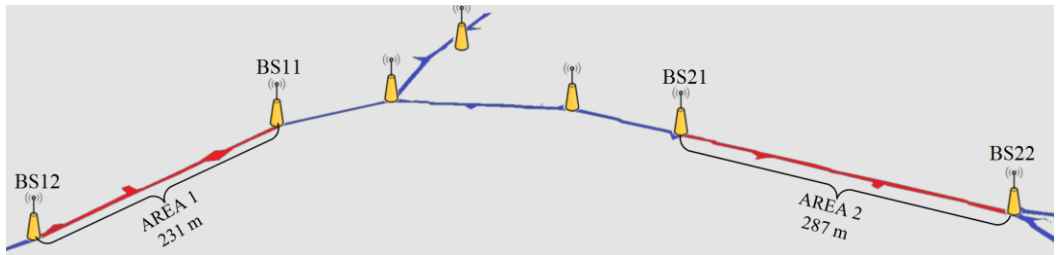

*Figure 7 The areas for the experiments (only areas in red are considered).*

The proposed approach was applied to the collected measurements. The measurements campaign was done in Area 1 and Area 2. The assessment procedure is similar to the corridor environment. Figures 8 and 9 demonstrate the measurements and the results of the estimations made by the proposed approach for Area 1 and Area 2 respectively. The accuracy of the distances is similar to the one shown in the corridor environment. Therefore, the graphical representation of the results is omitted for space saving reasons.

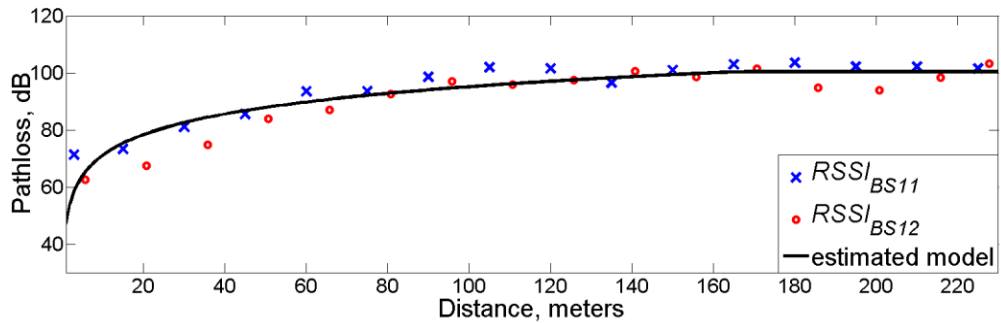

*Figure 8 Two-piece estimated model for measurements in Area1.*

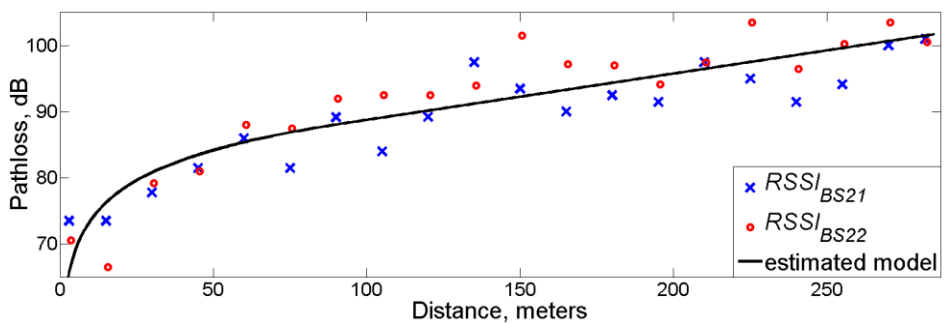

*Figure 9 Two-piece estimated model for measurements in Area2.*

The blue crosses in the figures denoted as $RSSI_{BS11}$ or $RSSI_{BS21}$ in the legends correspond to the measurements taken at the base stations at the beginning of the stretches. Red circles, denoted as $RSSI_{BS12}$ or $RSSI_{BS22}$, reflect the RSSI values measured by the base stations at the far ends of the areas. The figures show that in both cases the models with the estimated parameters fit to the provided measurements.

**Discussion and Conclusion**

This article proposed a simple practical approach for the self-adaptive estimation of the parameters for the two-piece path loss model. The estimated parameters accurately reflect the signal attenuation in the particular indoor or underground monitored environment. It is proposed to build an accurate Wi-Fi based positioning system using the simple two-piece path loss model with the estimated parameters. The proposed approach is self-adaptive in a sense that it does not require manual engineering of the model parameters. Instead, using the data from heterogeneous wireless technologies the model's parameters converge to the environment-specific values in several iterations. The accuracy of the proposed approach obviously increases as the history of RSSI measurements grow, i.e. with the number of estimation iterations.

Along the side with the validation process a procedure for placing the anchor points was developed. Namely, the quantity of the anchor points itself plays secondary role compared to the actual placement of RFID anchors. It is important to insert anchor points both in the presumed *near* and *far* regions of the two-piece model. In this case even very few anchor points per the monitored region are sufficient for accurate estimation of the path loss parameters.

On the level of a discussion the larger number of anchor points could be used to create a piecewise linear model of path loss on each interval between the RFID anchors. The development of this idea as well as its accuracy evaluation are subjects for future work. Another future work item concerns recognizing patterns of objects' mobility in the underground tunnels. Also particular resources should be dedicated for studies on how the effects such as shadow fading could affect the resulted models. Besides, the presence of vehicles obviously introduces additional impairments to radio signal propagation. On the other hand the information about the type of the moving object which is logged in the central control and supervision system could be overlaid with the RSSI measurements for further improvement of the positioning accuracy.

## Acknowledgements
This work is supported by the Swedish Governmental Agency for Innovation Systems (VINNOVA), under project 2013-00265 Industrial Positioning System. We also thank Mobilaris company for the assistance while performing measurements in the mine.

## Conflict of Interests
The authors declare that there is no conflict of interests regarding the publication of this paper.